\documentclass[oneside,11pt,a4paper,fleqn]{article}
\usepackage[T1]{fontenc}
\usepackage[utf8]{inputenc}
\usepackage{amsmath}
\usepackage{amssymb}
\usepackage{mathrsfs}
\usepackage{mathtools}
\usepackage{lmodern}
\usepackage{microtype}
\usepackage{contour}\contourlength{1pt}

\usepackage[hmarginratio=1:1,top=32mm,left=25mm,columnsep=20pt]{geometry} 
\usepackage[hang, small,labelfont=bf,up,textfont=it,up]{caption}
\usepackage{booktabs}

\definecolor{darkgreen}{rgb}{0.0, 0.6, 0.2}

\usepackage{paralist}
\usepackage{bbm}
\usepackage{tikz-cd} 
\usetikzlibrary{patterns,shapes,arrows}

\usepackage{dsfont}
\usepackage{braket}
\usepackage{nicefrac}

\usepackage{titlesec}
\usepackage{etoolbox}
\makeatletter
\patchcmd{\ttlh@hang}{\parindent\z@}{\parindent\z@\leavevmode}{}{}
\patchcmd{\ttlh@hang}{\noindent}{}{}{}
\makeatother

\usepackage{abstract} 

\advance \headheight by 3.0truept       

\usepackage[pdftitle={Discrete remnants of orbifolding},pdfauthor={Biermann, M\"utter, Parr, Ratz, Vaudrevange}, hypertexnames=true]{hyperref}
\hypersetup{bookmarksnumbered,colorlinks,
    linkcolor={black},
    citecolor={black},
    urlcolor={black}}

\usepackage{fancyhdr}
\usepackage{cleveref}

\DeclareMathOperator{\diag}{diag}
\newcommand{\I}{\mathrm{i}}

\newcommand{\Id}{\mathbbm{1}}

\newcommand{\Z}[1]{\ensuremath{\mathbbm{Z}_{#1}}}
\newcommand{\rep}[1]{\ensuremath{\op{#1}}}
\newcommand{\crep}[1]{\ensuremath{\op{\overline{#1}}}}

\newcommand{\su}[1]{\ensuremath{\mathfrak{su}(#1)}}
\newcommand{\so}[1]{\ensuremath{\mathfrak{so}(#1)}}

\newcommand{\SU}[1]{\ensuremath{\mathrm{SU}(#1)}}
\newcommand{\U}[1]{\ensuremath{\mathrm{U}(#1)}}
\newcommand{\SO}[1]{\ensuremath{\mathrm{SO}(#1)}}

\newcommand{\TCW}[1]{\ensuremath{{\mathsf{T}_{#1}^{(\mathrm{CW})}}}}
\newcommand{\op}[1]{\ensuremath{\boldsymbol{#1}}}

\title{Discrete remnants of orbifolding}

\begin{document}

\begin{titlepage}
\vspace*{-1cm}
\begin{flushright}
TUM-HEP 1206/19\\
UCI-2019-15\\
\end{flushright}

\vspace*{0.5cm}

\begin{center}
{\Huge\textbf{Discrete remnants of orbifolding}}

\vspace{0.7cm}

\textbf{Steffen Biermann$^{\,a}${},
Andreas M\"utter$^{\,b}${},
Erik Parr$^{\,b}${},\\
Michael Ratz$^{\,c}${},
Patrick K.S. Vaudrevange$^{\,b}${}
}
\\[8mm]
\textit{$^a$\small
~School of Mathematical Sciences, University of Nottingham, \\
Nottingham, NG7 2RD, United Kingdom}
\\[5mm]
\textit{$^b$\small
~Physik Department T75, Technische Universit\"at M\"unchen, \\
James--Franck--Stra\ss e, 85748 Garching, Germany}
\\[5mm]
\textit{$^c$\small
~Department of Physics and Astronomy, University of California,\\
Irvine, California 92697--4575, USA
}
\end{center}

\vspace{0.4cm}

\date{Version \today}

\begin{abstract}
We revisit the residual symmetries that survive the orbifold projections, and
find additional transformations that have been overlooked in the past. Some of
these transformations are outer automorphisms of the downstairs continuous
symmetry group. Examples for these transformations include the left--right
parity of the Pati--Salam model and its left--right symmetric subgroup. 
\end{abstract}

\end{titlepage}

\section{Introduction}

Gauge symmetry breaking via orbifolding
\cite{Dixon:1985jw,Dixon:1986jc,Ibanez:1987sn} is a popular alternative to
spontaneous breakdown of gauge symmetry in four dimensions. This has many
reasons, including the observation that the infamous doublet--triplet splitting
problem has a simple solution
\cite{Kawamura:1999nj,Barbieri:2000vh,Kawamura:2000ev,Altarelli:2001qj,Hebecker:2001wq,Hebecker:2001jb,Asaka:2001eh}.
The low--energy \emph{continuous} gauge symmetry in these models is well studied
\cite{Dixon:1986jc,Hebecker:2001jb}. The main purpose of this Letter is to
point out that there are additional discrete symmetries that have not been
identified, nor discussed, in this context thus far.

This Letter is organized as follows. In \Cref{sec:OrbifoldBasics} we review some
basic facts on orbifolding. In \Cref{sec:ResidualGaugeSymmetries} we revisit the 
conditions for residual symmetries and shall show that in the past some symmetries 
were missed. We illustrate this important fact by a few examples in 
\Cref{sec:Examples}, i.e.\ we give one example where a higher--dimensional 
$\SO{10}$ GUT is broken by an orbifold to Pati--Salam including left--right 
parity (a.k.a.\ $D$--parity). In addition, we present two examples 
which could be of relevance for flavor model--building from orbifold GUTs.
Finally, \Cref{sec:Summary} contains our summary. Some details are deferred to the 
appendices.

\section{Orbifold GUT breaking}
\label{sec:OrbifoldBasics}

Let us collect some basic facts on orbifolding. For the sake of definiteness 
we consider six--dimensional settings in which two dimensions get compactified,
but our findings do not depend on the number of dimensions. Consider a 
six--dimensional Yang--Mills theory with upstairs gauge group $\mathcal{G}$,
where we denote the generators of the Lie algebra in the Cartan--Weyl basis
$H_I$ and $E_w$ collectively by $\TCW{a}$. In a first step, this theory is
compactified on a two--torus $\mathbbm{T}^2$ defined by the lattice vectors 
$e_1$ and $e_2$, see \Cref{app:TorusSymmetries} for more details. We can 
choose the torus--lattice such that it exhibits a \Z{N} rotational symmetry 
$\vartheta$ with $\vartheta^N = \Id$, where for $N=3,4,6$ (i.e.\ the allowed 
orders $N\neq 2$ of the wall--paper groups in two dimensions) we set 
$\vartheta\,e_1 = e_2$, while in the case $N=2$ the basis vectors $e_1$ and 
$e_2$ have to be linear independent. In order to orbifold the two--torus 
$\mathbbm{T}^2$ to a $\mathbbm{T}^2/\Z{N}$ orbifold we mod out this $\Z{N}$ 
symmetry, i.e.\ we identify points $y$ on $\mathbbm{T}^2$ which are related 
by a $(360/N)^\circ$ rotation,
\begin{equation}\label{eq:ZNgeom}
 y ~\xmapsto{~\Z{N}~}~ \vartheta\,y ~\sim~ y\;.
\end{equation}
Note that under this geometrical action our six--dimensional fields transform as
\begin{equation}
 V^\mu(x,y) ~\xmapsto{~\Z{N}~}~ V^\mu(x, \vartheta^{-1}\, y)\;, \quad\mathrm{and}\quad \chi(x,y) ~\xmapsto{~\Z{N}~}~  \exp\left(\frac{2\pi\I}{N}\right)\, \chi(x, \vartheta^{-1}\, y)\;,
\end{equation}
where the $\chi$ fields transform as the internal components of a 6D vector 
$V^M(x,y)$ of six--dimensional Lorentz symmetry.
Moreover, the \Z{N} orbifold can be extended from its pure geometric action
\Cref{eq:ZNgeom}  to include a discrete $\Z{N}$ transformation from the gauge
symmetry $\mathcal{G}$, i.e.
\begin{equation}\label{eqn:orbifoldactionP}
\TCW{a} ~\xmapsto{~\Z{N}^\mathrm{orb.}~}~ P\,\TCW{a}\,P^{-1}  \quad\mathrm{with}\quad P^N ~=~ \Id\;,
\end{equation}
where $P \in \mathcal{G}$ acts as a discrete gauge transformation\footnote{We
ignore the  possibility to choose an outer automorphism of $\mathcal{G}$ as
gauge action \cite{Hebecker:2001jb}.  Furthermore, the order of $P$ can in 
general differ from the order of $\vartheta$.}, see
\Cref{eq:GaugeTransformation} with  $U=P=\mathrm{constant}$. Since we restrict
ourselves to Abelian  orbifolds, we can choose the Cartan generators $H_I$
of $\mathcal{G}$ such that $P$ can be expanded as
\begin{equation}\label{eq:PInTermsOfH}
 P ~=~ \exp\left(2\pi\I\, V\cdot H\right)\;,
\end{equation}
where the vector $V$ is ``quantized'' such that $P^N = \Id$.

\paragraph{Orbifold conditions.} Next, in addition to the torus boundary
conditions~\eqref{eq:BoundaryConditionsTorus}, we impose orbifold boundary
conditions
\begin{subequations}\label{eq:BoundaryConditionsOrbifold}
\begin{align}
V^\mu(x, \vartheta\, y) &~= ~ P\,V^\mu(x,y)\,P^{-1}\;, \\
\chi(x, \vartheta\, y)  &~=~ \exp\left(\frac{2\pi\I}{N}\right)\, P\,\chi(x,y)\,P^{-1}\;.
\end{align}
\end{subequations}
Using
\begin{equation}\label{eq:TrafoOfCWGenerators}
P\,H_I\,P^{-1} ~=~ H_I \quad\mathrm{and}\quad P\,E_w\,P^{-1} ~=~ \exp\left(2\pi\I\, V\cdot w\right)\, E_w\;,
\end{equation} 
where $w$ denotes the root vector of $E_w$, we obtain
\begin{subequations}\label{eq:BoundaryConditionsOrbifoldDiag}
\begin{align}
V^\mu_I(x, \vartheta\, y) &~=~ V^\mu_I(x,y)\;, \\
V^\mu_w(x, \vartheta\, y) &~=~ \exp\left(2\pi\I\, V\cdot w\right)\, V^\mu_w(x,y)\;, \\
\chi_I(x, \vartheta\, y)  &~=~ \chi_I(x,y)\;, \\
\chi_w(x, \vartheta\, y)  &~=~ \exp\left(2\pi\I\, \left(V\cdot w + \frac{1}{N}\right)\right)\, \chi_w(x,y)\;.
\end{align}
\end{subequations}

\section{Residual gauge symmetries}
\label{sec:ResidualGaugeSymmetries}

We consider the possibility of unbroken discrete symmetries from $\mathcal{G}$. In this case, a 
symmetry transformation from $\mathcal{G}$ remains unbroken if it commutes with the orbifold boundary 
condition~\eqref{eq:BoundaryConditionsOrbifold}, i.e.
\begin{equation}\label{eqn:UnbrokenDiscreteSymmetries}
\begin{tikzcd}
V^\mu_a(x,y)\,\TCW{a} \arrow[mapsto]{dd}{\mathcal{G}} \arrow[mapsto]{r}{O}
& 
V^\mu_a(x,\vartheta^{-1}\,y)\, P\,\TCW{a}\,P^{-1} \arrow[mapsto]{d}{\mathcal{G}} 
\\
& 
V^\mu_a(x,\vartheta^{-1}\,y)\, U\,P\,\TCW{a}\,P^{-1}\,U^{-1} 
\arrow[color=white]{d}[black]{\rotatebox{270}{\Large$\stackrel{!}{=}$}} 
\\ 
V^\mu_a(x,y)\,U\,\TCW{a}\,U^{-1} \arrow[mapsto]{r}{O} & V^\mu_a(x,\vartheta^{-1}\,y)\, P\,U\,\TCW{a}\,U^{-1}\,P^{-1} \;,
\end{tikzcd}
\end{equation}
for a global, unbroken transformation $U \in \mathcal{G}$, see \Cref{eq:GaugeTransformation}. 
Consequently, we obtain the condition
\begin{equation}
\TCW{a}\, \left(P^{-1}\,U^{-1}\,P\,U\right) ~=~ \left(P^{-1}\,U^{-1}\,P\,U\right)\, \TCW{a}\;.
\end{equation}
Due to Schur's lemma, it follows that $P^{-1}\,U^{-1}\,P\,U \propto \Id$.
Furthermore, $P$ is of  order $N$ (i.e.\ $P^N = \Id$) yielding our main
condition for unbroken symmetries after orbifolding
\begin{equation}\label{eq:DiscreteFromGauge}
P^{-1}U^{-1}P\,U ~=:~ [P, U] ~=~\omega^k\, \Id \qquad\mathrm{for}\quad  k ~\in~\{0,1,\ldots,N-1\}\;,
\end{equation}
where $\omega = \mathrm{e}^{\frac{2\pi\I}{N}}$ and we use the definition of the 
(grouptheoretical) commutator of two \emph{group} elements (as opposed to
Lie algebra elements), $[A,B]=A^{-1} B^{-1}A\,B$~\cite{Ramond:2010zz}. Since 
$P, U \in \mathcal{G}$ also $[P, U]$ must be from $\mathcal{G}$. Moreover, 
$[P, U] \propto \Id$. Thus, $[P, U]$ must be from the center of $\mathcal{G}$, 
i.e.\
\begin{equation}
\omega^k\, \Id ~\in~ Z(\mathcal{G})\qquad\mathrm{for\ some}\quad k ~\in~\{0,1,\ldots,N-1\}\;.
\end{equation}
This constrains the allowed values of $k$. For example, the center of $\SU{M}$
is $\Z{M}$, while  $\omega$ is of order $N$. That is, these additional
residual symmetries require the order of the orbifold twist and the dimension of
the group center to be not coprime.

\subsection{\boldmath Unbroken continuous gauge symmetries \unboldmath}

There are two related ways to identify the unbroken gauge symmetries after
orbifolding.

First, as is well known, the unbroken gauge interactions are mediated by the
zero--modes of the gauge bosons. These are the modes with trivial boundary
conditions \Cref{eq:BoundaryConditionsOrbifoldDiag}. Thus, the gauge bosons 
$V^\mu_I(x,y)$ and $V^\mu_w(x,y)$, which are associated to the Cartan generators
$H_I$  and to those roots $w$ for which $V\cdot w\in\mathbbm{Z}$, have trivial
boundary conditions and hence massless modes in four dimensions. 

Second, we can use our main condition~\eqref{eq:DiscreteFromGauge} to  identify
the unbroken continuous symmetries \cite{Hebecker:2001jb}. The unbroken
continuous symmetries are continuously connected to the  identity $U = \Id$.
Hence, we have to set $k=0$ in  \Cref{eq:DiscreteFromGauge} and expand  $U =
\exp\left(\I\,\alpha_a\,\TCW{a}\right) \approx \Id + \I\,\alpha_a\,\TCW{a}$.  In
this way, \Cref{eq:DiscreteFromGauge} yields the condition for a  generator of
the unbroken gauge symmetry
\begin{equation}\label{eq:UnbrokenGaugeSymmetry}
 P^{-1} \left(\alpha_a\,\TCW{a}\right)\, P ~=~ \left(\alpha_a\,\TCW{a}\right)\;.
\end{equation}
Since the boundary condition $P$ is expanded in terms of the Cartan  generators
$H_I$, \Cref{eq:PInTermsOfH}, we can use \Cref{eq:TrafoOfCWGenerators}  to
confirm that the Cartan generators $H_I$ and the generators $E_w$ with  $V\cdot
w\in\mathbbm{Z}$ satisfy \Cref{eq:UnbrokenGaugeSymmetry}, i.e.\  they remain
unbroken after orbifolding.

\subsection{\boldmath Unbroken discrete gauge symmetries\unboldmath}

In addition to the unbroken continuous gauge symmetries, our main 
condition~\eqref{eq:DiscreteFromGauge} can have additional solutions which 
then lead to further discrete remnants from the higher--dimensional gauge 
symmetry $\mathcal{G}$.  Importantly, these discrete symmetries can originate 
from our main  condition~\eqref{eq:DiscreteFromGauge} either for $k=0$ (see the 
example in  \Cref{sec:PatiSalam}) or for $k\neq 0$ (see the examples in 
\Cref{sec:kNotZero}).

\section{Examples and applications}
\label{sec:Examples}

In this section, we illustrate our general findings in a few examples.

\subsection[Gauge origin of D-parity and left-right parity]{\boldmath Gauge origin of $D$--parity and left--right parity}
\label{sec:PatiSalam}

The Pati--Salam model~\cite{Pati:1974yy}can have, in addition to the continuous
gauge group 
\begin{equation}
 G_\mathrm{PS}~=~\SU4\times\SU2_\mathrm{L}\times\SU2_\mathrm{R}\;,
\end{equation}
a \Z2 symmetry $D$ that exchanges the \SU2 factors and acts on \SU4
representations as complex conjugation. This symmetry is part of the \SO{10}
supergroup containing $G_\mathrm{PS}$, and can be preserved in 4D models of
grand unification if one breaks \SO{10} by giving a VEV to a \rep{54}--plet 
\cite{Kibble:1982dd,Chang:1983fu}. At the level of the left--right symmetric
subgroup of $G_\mathrm{PS}$,
$G_\mathrm{LR}=\SU3_\mathrm{C}\times\SU2_\mathrm{L}\times\SU2_\mathrm{R}\times\U1_{B\!-\!L}$,
this \Z2 is the well--known left--right parity~\cite{Mohapatra:1979ia}. That is,
the symmetries of these settings are
\begin{equation}
 \left[\SU4\times\SU2_\mathrm{L}\times\SU2_\mathrm{R}\right]\rtimes\Z2
 ~\mathrm{or}~
 \left[\SU3_\mathrm{C}\times\SU2_\mathrm{L}\times\SU2_\mathrm{R}\times\U1_{B\!-\!L}\right]\rtimes\Z2
 \;.
\end{equation} 
The purpose of the following discussion is to show that this \Z2 factor is a
residual symmetry of the corresponding orbifold GUT, which to our knowledge has
not been pointed out before.

To this end, consider a theory with \SO{10} symmetry in higher dimensions 
compactified on a \Z2 orbifold such as $\mathbbm{S}^1/\Z2$ or 
$\mathbbm{T}^2/\Z{2}$. We choose the GUT breaking boundary condition
\begin{equation}
P_\mathrm{PS}~=~\diag(-\Id_6;\Id_4)\;.
\end{equation}
As is well known, the continuous low--energy gauge symmetry is
$G_\mathrm{PS}$~\cite{Asaka:2001eh}. However, there is an additional \Z2
symmetry.

In more detail, our main condition~\eqref{eq:DiscreteFromGauge} yields
\begin{equation}\label{eq:PSOrbifoldBreaking}
[P_\mathrm{PS}, U_{(k)}] ~=~ (-1)^k\, \Id \qquad\mathrm{for}\quad  k ~\in~ \{0,1\}\;,
\end{equation}
and we search for the unbroken elements $U_{(k)} \in \SO{10}$. For $k=0$ 
condition~\eqref{eq:PSOrbifoldBreaking} reads
\begin{equation}
P_\mathrm{PS}\, U_{(0)} ~=~ U_{(0)}\; P_\mathrm{PS}\;.
\end{equation}
The most general $\SO{10}$ matrix satisfying this condition reads
\begin{equation}
U_{(0)} ~=~ \begin{pmatrix} O_6 &0 \\
  0 & O_4
 \end{pmatrix}~\in~\SO{10}\;.
\end{equation}
Consequently, we find the conditions
\begin{equation}
O_6^T\, O_6 ~=~ \Id_6 \quad\mathrm{and}\quad O_4^T\, O_4 ~=~ \Id_4
\quad\mathrm{and}\quad \det O_6 ~=~ \det O_4 ~=~ \pm 1\;.
\end{equation}
Hence, $U_{(0)}$ with $\det O_6 = \det O_4 = +1$ yields 
\begin{equation}
O_6 ~\in~ \SO{6} ~\simeq~ \SU{4} \quad\mathrm{and}\quad O_4 ~\in~ \SO{4} ~\simeq~ \SU{2}_\mathrm{L}\times\SU{2}_\mathrm{R}\;,
\end{equation}
while $U_{(0)}$ with $\det O_6 = \det O_4 = -1$ can be generated by
\begin{equation}\label{eq:PSDefinitionOfZ2}
O_6~=~\diag(1,1,1,1,1,-1)\,O_6' \quad\mathrm{and}\quad 
O_4 ~=~\diag(1,1,1,-1)\, O_4'\;,
\end{equation}
for $O_6'\in\SO{6}\simeq\SU4$ and
$O_4'\in\SO{4}\simeq\SU2_\mathrm{L}\times\SU2_\mathrm{R}$.\footnote{Note that
the ``$\simeq$'' means ``up to \Z2 factors'', but these \Z2's are different from
the one we are going to discuss next.} Let us remark that setting $k=1$ in our main 
condition~\eqref{eq:PSOrbifoldBreaking} does not yield further unbroken symmetries.

Consequently, the $\Z{2}$ orbifold boundary condition $P_\mathrm{PS}$ breaks
$\SO{10}$ to 
\begin{equation}
G_\mathrm{PS} ~=~ \left(\SU4\times\SU2_\mathrm{L}\times\SU2_\mathrm{R}\right) \rtimes \Z{2}\;,
\end{equation}
where the generator of the additional $\Z{2}$ remnant symmetry can be chosen
to be
\begin{equation}\label{eq:Dparity}
D~=~\diag(-1,1,1,1,1,1; 1,-1,-1,-1)\;.
\end{equation}
Here, we write $D$ in this suggestive way because this will make it very obvious
how this \Z2 acts. We could have represented it by any diagonal matrix with
entries $\pm1$  subject to the condition that the number of $-1$ on either
sides of the semicolon is odd. 

How does this \Z2 act on representations? Consider first the \SO4 subblock.
There, the transformation $D$ can be understood by analogy to parity 
acting on spinors $(\nicefrac{1}{2},0) \oplus (0,\nicefrac{1}{2})$ of 
$\SU2\times\SU2$ in 4D Euclidean space--time: parity interchanges the \SU2 
representations. Translated to Pati--Salam, $D$ acts on 
$(\rep{r}_\mathrm{L},\rep{r}_\mathrm{R})$ of $\SU2_\mathrm{L}\times\SU2_\mathrm{R}$ as
\begin{equation}
 (\rep{r}_\mathrm{L},\rep{r}_\mathrm{R})~\xmapsto{~D~}~
 (\rep{r}_\mathrm{R},\rep{r}_\mathrm{L})\;,
\end{equation}
see also \Cref{app:LRParity} for an explicit computation how $D$ acts on 
$\SU2_\mathrm{L}\times\SU2_\mathrm{R}$. Similarly, $D$ acts on the 
$\SO6\simeq\SU4$ subgroup in analogy to (an Euclidean version of)  time
reversal, so for any \SU4 representation $\rep{r}_4$
\begin{equation}
 \rep{r}_4~\xmapsto{~D~}~\crep{r}_4\;.
\end{equation}
Altogether a representation $(\rep{r}_4,\rep{r}_\mathrm{L},\rep{r}_\mathrm{R})$ of  
$\SU4\times\SU2_\mathrm{L}\times\SU2_\mathrm{R}$ transforms under $D$ as
\begin{equation}
 (\rep{r}_4,\rep{r}_\mathrm{L},\rep{r}_\mathrm{R})~\xmapsto{~D~}~
 (\crep{r}_4,\rep{r}_\mathrm{R},\rep{r}_\mathrm{L})\;.
\end{equation}
So this \Z2 exchanges $(\rep{4},\rep{2},\rep{1})$ and
$(\crep{4},\rep{1},\rep{2})$, i.e.\ the left- and right--handed fermions of the
standard model. That is, this simple orbifold GUT gives rise to the well--known
left--right parity~\cite{Mohapatra:1979ia}, where it originates from \SO{10} and
is hence clearly a discrete gauge symmetry. Ironically, the representation of
its generator \eqref{eq:Dparity} supports the naming in \cite{Mohapatra:1979ia},
where this transformation has been called parity. Even though it is not the
ordinary space--time transformation that gets broken spontaneously there, as the
left--right symmetric model is chiral and even in its unbroken phase does not
preserve parity, this transformation does act on the $\SO{6}\simeq\SU4$ and
$\SO{4}\simeq\SU2_\mathrm{L}\times\SU2_\mathrm{R}$ representations in an
analogous way as space--time parity does.

Altogether we have found that the breaking pattern of the \SO{10} orbifold GUT
is
\begin{equation}
 \SO{10}~\xrightarrow{\Z2~\mathrm{orbifold}}~
 \left[\SU4\times\SU2_\mathrm{L}\times\SU2_\mathrm{R}\right]\rtimes\Z2\;,
\end{equation}
where the \Z2 corresponds to the left--right parity and is in particular a
nontrivial outer automorphism of $G_\mathrm{PS}$. It is amusing to see that the
same mechanism that breaks the gauge symmetry and provides us with a solution to
the doublet--triplet splitting problem automatically leads to this symmetry.

\begin{figure}[t]
 \centering
 \begin{tikzpicture}
  \begin{scope}[local bounding box=so5]
   \draw foreach \X in {0,...,4}
    {foreach \Y in {0,...,4}
    {(\X,\Y) circle[radius=0.4mm] 
    ({\X-1/2},{\Y+1/2}) circle[radius=0.4mm] }};
   \draw[fill=red] (2,3) circle[radius=0.5mm];
   \draw[fill=red] (2,1) circle[radius=0.5mm];
   \draw[fill=red] (3,2) circle[radius=0.5mm];
   \draw[fill=red] (1,2) circle[radius=0.5mm];
   \draw[fill=red] (2+1/2,2+1/2) circle[radius=0.5mm];
   \draw[fill=red] (1+1/2,2+1/2) circle[radius=0.5mm];
   \draw[fill=red] (2+1/2,1+1/2) circle[radius=0.5mm];
   \draw[fill=red] (1+1/2,1+1/2) circle[radius=0.5mm];
   \draw[black,thick,-stealth] (2,2) -- (3,2)
   	node[below,black]{\contour{white}{$\alpha_{(1)}^{\mathfrak{so}(5)}$}};
   \draw[black,thick,-stealth] (2,2) -- (1.5,2.5) 
   	node[left,black]{\contour{white}{$\alpha_{(2)}^{\mathfrak{so}(5)}$}};
   \draw[dashed,green!60!black] (-0.5,-0.5) -- (4.5,4.5);
  \end{scope}
  \node[above,black] at (so5.north){$\mathfrak{so}(5)$};
  \begin{scope}[local bounding box=so4,xshift=6cm]
   \draw foreach \X in {0,...,4}
    {foreach \Y in {0,...,4}
    {(\X,\Y) circle[radius=0.4mm]}};
   \draw[fill=red] (2,3) circle[radius=0.5mm];
   \draw[fill=red] (2,1) circle[radius=0.5mm];
   \draw[fill=red] (3,2) circle[radius=0.5mm];
   \draw[fill=red] (1,2) circle[radius=0.5mm];
   \draw[black,thick,-stealth] (2,2) -- (3,2)
   node[below,black]{\contour{white}{$\alpha_{(1)}^{\mathfrak{su}(2)_\mathrm{L}}$}};
   \draw[black,thick,-stealth] (2,2) -- (2,3)
   node[left,black]{\contour{white}{$\alpha_{(1)}^{\mathfrak{su}(2)_\mathrm{R}}$}};
   \draw[dashed,green!60!black] (-0.5,-0.5) -- (4.5,4.5);
   \draw[green!60!black,ultra thick,latex-latex,shorten >=1mm,shorten <=1mm] 
    (3,2) to[bend right=20] node[midway,above,sloped]{\contour{white}{$D$}} (2,3);
  \end{scope} 
  \node[above,black] at (so4.north){$\mathfrak{so}(4)=\mathfrak{su}(2)_\mathrm{L}\oplus\mathfrak{su}(2)_\mathrm{R}$};
 \end{tikzpicture}
 \caption{Root lattices of $\mathfrak{so}(5)$ and its
  $\mathfrak{so}(4)=\mathfrak{su}(2)_\mathrm{L}\oplus\mathfrak{su}(2)_\mathrm{R}$ subalgebra. 
  The roots of the respective algebra are depicted by filled circles and the simple roots 
  are given by arrows. A Weyl reflection w.r.t.\ the plane perpendicular to 
  $\alpha_{(2)}^{\mathfrak{so}(5)}$ exchanges the generators of the two $\mathfrak{su}(2)$ algebras. 
  Hence, the outer automorphism $D$ of $\mathfrak{su}(2)_\mathrm{L}\oplus\mathfrak{su}(2)_\mathrm{R}$ 
  is generated by the Weyl reflection at the broken root $\alpha_{(2)}^{\mathfrak{so}(5)}$ of 
  $\mathfrak{so}(5)$.}
 \label{fig:WeylSO5toSO4}
\end{figure}
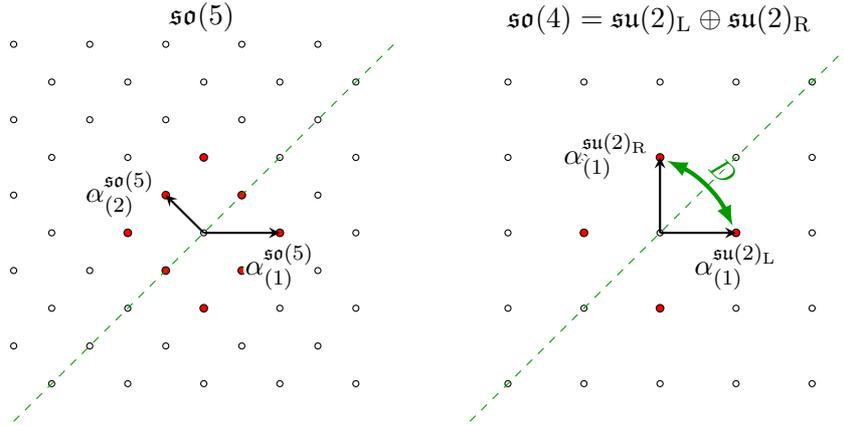

This parity has a simple geometric interpretation in terms of root lattices, 
which already can be obtained from a lower--dimensional example. Consider the 
breaking of \SO{5} to \SO{4} with a twist $P_5=\diag(1,-1,-1,-1,-1)\in\SO{5}$. 
This breaking removes a simple root from the root lattice (see 
\Cref{fig:WeylSO5toSO4}), and the simple roots of 
$\mathfrak{su}(2)_\mathrm{L}\oplus\mathfrak{su}(2)_\mathrm{R}$
span a sublattice of the original $\mathfrak{so}(5)$ lattice. However, the Weyl
reflection w.r.t.\ the plane orthogonal to the ``broken'' root
$\alpha_{(2)}^{\mathfrak{so}(5)}$ is a symmetry of the 
$\mathfrak{su}(2)_\mathrm{L}\oplus\mathfrak{su}(2)_\mathrm{R}$ sublattice, and exchanges
(the generators of) the $\mathfrak{su}(2)$ algebras. 

The analogous statement holds in the full Pati--Salam example, but depicting
the transformation $D$ as a Weyl reflection is more difficult since the rank of
$\mathfrak{so}(10)$ is 5. As we shall see, the residual transformations in the
examples in \Cref{sec:kNotZero} can also be related to elements of the Weyl group. 

Discussing the phenomenological implications of this symmetry is beyond the
scope of this work, we only note the revived interest in this transformation in
\cite{Hall:2019qwx} and references therein.

\subsection{Non--Abelian residual symmetries}
\label{sec:kNotZero}

In what follows, we present two examples in which the higher--dimensional 
gauge group gets broken by the orbifold to a semi--direct product of an Abelian 
gauge symmetry with a discrete $\Z{N}$ factor. Such symmetries naturally 
contain non--Abelian discrete groups that can be used as flavor symmetries.

\subsubsection{\boldmath $\mathbbm{T}^2/\Z{4}$ Orbifold GUT\unboldmath}

We choose a six--dimensional gauge symmetry $\mathcal{G}=\SU{2}$ and $|e_1|=|e_2|$ with 
$e_1\cdot e_2 = 0$. This lattice has a $\Z{4}$ rotational symmetry $\vartheta$ that we divide out in 
order to construct a $\mathbbm{T}^2/\Z{4}$ orbifold. The associated gauge embedding $P$ is chosen 
as
\begin{equation}
P ~=~ \begin{pmatrix}
\I &  0  \\
 0 &-\I  \\
\end{pmatrix} ~\in~ \SU{2} \quad\mathrm{where}\quad P^4 ~=~ \Id\;.
\end{equation}
Then, the unbroken symmetry is given by those $U_{(k)} \in \SU{2}$ that satisfy
\begin{equation}\label{eq:Z4UnbrokenSymmetry}
[P,U_{(k)}] ~=~ \exp\left(\frac{2\pi\I\, k}{4}\right) \Id \quad\mathrm{where}\quad k ~\in~ \{0,1,2,3\}\;.
\end{equation}
Since $P, U_{(k)} \in \SU{2}$, the right-hand side of \Cref{eq:Z4UnbrokenSymmetry} has to be 
an element of $\SU{2}$, too. Moreover $[P,U_{(k)}] \propto \Id$, thus, it has to be from the center 
$Z(\SU{2}) = \Z{2}$. Consequently, \Cref{eq:Z4UnbrokenSymmetry} can only have solutions for 
$k\in\{0,2\}$.

To find all solutions of \Cref{eq:Z4UnbrokenSymmetry} we parameterize a general element 
$U_{(k)}\in\SU{2}$ using $p,q\in\mathbbm{C}$ as
\begin{equation}
U_{(k)} ~=~ \begin{pmatrix}
 p       &  q  \\
-\bar{q} & \bar{p} \\
 \end{pmatrix} ~\in~ \SU{2} 
 \quad\mathrm{where}\quad 
 \det(U_{(k)}) = |p|^2 + |q|^2 = 1\;.\label{eq:U(k)}
\end{equation}
Then, \Cref{eq:Z4UnbrokenSymmetry} reads
\begin{equation}\label{eq:Z4UnbrokenSymmetry2}
[P,U_{(k)}] ~=~ \begin{pmatrix}
 |p|^2-|q|^2 & 2\bar{p}q  \\
-2p\bar{q}   & |p|^2-|q|^2 \\
\end{pmatrix} ~\stackrel{!}{=}~ \exp\left(\frac{2\pi\I\, k}{4}\right) \Id\;,
\end{equation}
which is equivalent to
\begin{equation}\label{eq:Z4UnbrokenSymmetry3}
|p|^2-|q|^2 ~\stackrel{!}{=}~ \exp\left(\frac{2\pi\I\, k}{4}\right) \quad\mathrm{and}\quad \bar{p}q ~\stackrel{!}{=}~ 0\;.
\end{equation}
Now, since $|p|^2-|q|^2 \in \mathbbm{R}$ we see explicitly that \Cref{eq:Z4UnbrokenSymmetry2} 
has no solutions for $k\in\{1,3\}$.

Setting $k=0$ in \Cref{eq:Z4UnbrokenSymmetry3} we find the unbroken gauge symmetry given by
$|p|^2 = 1$ (hence, $p=\mathrm{e}^{\I\alpha}$) and $q=0$, i.e.
\begin{equation}
U_{(0)} ~=~ U_{(0)}(\alpha) ~=~ \begin{pmatrix}
 \mathrm{e}^{\I\alpha} & 0  \\
 0            & \mathrm{e}^{-\I\alpha} \\
\end{pmatrix} ~\in~ \SU{2} \;,
\end{equation}
where $\alpha \in [0,2\pi)$. This yields an unbroken $\U{1}$ gauge symmetry. On the other hand, 
setting $k=2$ in \Cref{eq:Z4UnbrokenSymmetry3} yields $p=0$ and $|q|^2 = 1$ (thus, 
$q=\I\mathrm{e}^{\I\alpha}$, where the additional factor $\I$ has been introduced for later 
convenience), i.e.\
\begin{equation}
U_{(2)} ~=~ \begin{pmatrix} 
 0             & \I\mathrm{e}^{\I\alpha}  \\
\I\mathrm{e}^{-\I\alpha} & 0 \\
\end{pmatrix} ~=~ \begin{pmatrix}
 \mathrm{e}^{\I\alpha} & 0  \\
 0            & \mathrm{e}^{-\I\alpha} \\
\end{pmatrix}\, \begin{pmatrix}
 0 & \I  \\
\I & 0 \\
\end{pmatrix}~\in~ \SU{2}\;,\label{eq:U(1)/Z2:U(2)}
\end{equation}
where $\alpha \in [0,2\pi)$.

Consequently, the unbroken symmetry of $\SU{2}$ is generated by a 
$\U{1}$ and a $\Z{4}$, i.e.\
\begin{equation}
U_{(0)}(\alpha) ~=~ \begin{pmatrix}
 \mathrm{e}^{\I\alpha} & 0  \\
 0            & \mathrm{e}^{-\I\alpha} \\
\end{pmatrix} \quad\mathrm{and}\quad U_{(2)} ~=~ \begin{pmatrix}
 0 & \I  \\
\I & 0 \\
\end{pmatrix}\;,
\end{equation}
where $(U_{(2)})^2=-\Id = U_{(0)}(\pi) \in \U{1}$. The $\Z{4}$ 
transformation $U_{(2)}$ acts on the gauge bosons as $\Z{2}$, i.e.\
\begin{equation}\label{eq:UInSU2Example}
V^\mu_a(x,y)\,\TCW{a} ~\mapsto~ V^\mu_a(x,y)\,U_{(2)}\,\TCW{a}\,U_{(2)}^{-1}\;,
\end{equation}
see the diagram~\eqref{eqn:UnbrokenDiscreteSymmetries}. By explicitly 
choosing the Cartan--Weyl basis $H=\frac{1}{\sqrt{2}}\sigma_3$ and 
$E_\pm=\frac{1}{2}\left(\sigma_1\pm\I\sigma_2\right)$, one verifies that 
$U_{(2)}$ in \Cref{eq:UInSU2Example} can be understood as the action of 
the unbroken element $w$ of the Weyl group of $\mathfrak{su}(2)$, i.e.\
\begin{align}\label{eq:SU2WeylTrafo}
w~:~
\begin{pmatrix}
H\\E_+\\E_-
\end{pmatrix}
~\mapsto~
\begin{pmatrix}
-H\\E_-\\E_+
\end{pmatrix}\;.
\end{align}
In summary, the six--dimensional $\SU{2}$ gauge symmetry is broken by this
$\Z{4}$ orbifold according to
\begin{equation}\label{eqn:SU2BreakingPattern}
 \SU2 ~\xrightarrow{~\Z4^\mathrm{orb.}~}~ \left(\U1\rtimes\Z4\right)/\Z{2}\;.
\end{equation}
Let us remark that this unbroken symmetry after orbifolding contains, for
example,  the binary dihedral groups $Q_N$ with $N = \textrm{even}$ as 
subgroups~\cite{Ishimori:2010au}, including the quaternion group for $N=4$.

\subsubsection{\boldmath $\mathbbm{T}^2/\Z{3}$ Orbifold GUT\unboldmath}

Next, we choose a six--dimensional gauge symmetry $\mathcal{G}=\SU{3}$ and
$|e_1|=|e_2|$ with  $e_1\cdot e_2 = -\nicefrac{|e_1|^2}{2}$. This lattice has a
$\Z{3}$ rotational symmetry $\vartheta$ that we divide out in  order to
construct a $\mathbbm{T}^2/\Z{3}$ orbifold. The associated gauge embedding $P$
is chosen as
\begin{equation}
P ~=~ \begin{pmatrix}
\omega & 0  & 0\\
 0 & \omega^2 & 0\\
 0 &  0 & 1 
\end{pmatrix} ~\in~ \SU{3} \quad\mathrm{where}\quad P^3 ~=~ \Id\;,
\end{equation}
where $\omega = \exp \nicefrac{2\pi\I}{3}$. Then, the unbroken symmetry is given by those 
$U_{(k)} \in \SU{3}$ that satisfy 
\begin{equation}\label{eq:Z3UnbrokenSymmetry}
[P,U_{(k)}] ~=~ \exp\left(\frac{2\pi\I\, k}{3}\right) \Id \quad\mathrm{where}\quad k~\in~\{0,1,2\}\;.
\end{equation}
Since $P, U_{(k)} \in \SU{3}$, the right-hand side of \Cref{eq:Z3UnbrokenSymmetry} has to be 
an element of $\SU{3}$, too. Moreover $[P,U_{(k)}] \propto \Id$, thus, it has to be from the center 
$Z(\SU{3}) = \Z{3}$. Consequently, \Cref{eq:Z3UnbrokenSymmetry} can have solutions for all 
cases $k\in\{0,1,2\}$.

The unbroken symmetry can be generated by two $\U{1}$ factors
\begin{equation}
U_{(0)} ~=~ \begin{pmatrix}
 \mathrm{e}^{\I(\alpha+\beta)} & 0 & 0\\
 0 &  \mathrm{e}^{\I(\alpha-\beta)} & 0\\
 0 & 0 &  \mathrm{e}^{-2\I\alpha} 
\end{pmatrix} ~\in~ \SU{3}
\end{equation}
and two discrete transformations
\begin{equation}
U_{(1)} ~=~ \begin{pmatrix}
 0 & 0 & 1\\
 1 & 0 & 0\\
 0 & 1 & 0 
\end{pmatrix} ~\in~ \SU{3} \quad,\quad 
U_{(2)} ~=~ \begin{pmatrix} 
 0 & 1 & 0\\
 0 & 0 & 1\\
 1 & 0 & 0 
\end{pmatrix} ~\in~ \SU{3}\;,
\end{equation}
where $U_{(2)} = (U_{(1)})^2$. Since $(U_{(1)})^3 = \Id$, $U_{(1)}$ generates an unbroken \Z{3}.
Consequently, the six--dimensional $\SU{3}$ gauge symmetry is broken by the $\Z{3}$ orbifold 
according to (cf.\ \cite{Beye:2014nxa,Beye:2015wka})
\begin{equation}\label{eqn:SU3BreakingPattern}
\SU3 ~\xrightarrow{~\Z3^\mathrm{orb.}~}~ \Big[\U1\times\U1\Big]\rtimes\Z3\;.
\end{equation}
Again, the \Z3 can be understood as a remnant of the Weyl group: if we denote
the Weyl reflection w.r.t.~the root $\alpha$ by $w_\alpha$, conjugating with
$U_{(1)}$ has the same action on the generators as the Weyl transformation
$w_{\alpha_{(1)}}w_{\alpha_{(2)}}$, where $\alpha_{(I)}$, $I=1,2$, denote the
simple roots of $\SU{3}$. The \U1 factors emerge from the standard gauge
symmetry breaking by orbifold boundary conditions to  the commuting subgroup,
see for example \cite[Equation~(6)]{Hebecker:2001jb}. However, to our knowledge,
there is no systematic way in the previous literature how to derive the
(non--commuting) $\Z{3}$ factor. We also note that if one breaks the \U1 factors
down to \Z3 symmetries, this leaves us with $(\Z3\times\Z3)\rtimes\Z3$, which is
known as $\Delta(27)$ and has been proposed as a flavor symmetry.

\section{Summary}
\label{sec:Summary}

We discussed how gauged discrete symmetries emerge from orbifolds. Although we
used the field--theoretic constructions the discussion is purely
group--theoretical and applies to string--derived orbifolds as well. We identify
residual discrete symmetries. These include the so--called left--right parity of
the Pati--Salam model or its left--right symmetric subgroup, which, to the best
of our knowledge, have been overlooked in the literature so far. These
symmetries are inner automorphisms of the upstairs symmetry group but outer
automorphisms of the orbifolded setup. Notably, we find that these symmetries do
not have to commute with the orbifold twist. Rather, the transformations $U$
have to fulfill the weaker condition
\begin{align}
 P^{-1}U^{-1}P\,U ~=~ \omega^k\, \Id ~\in~Z(\mathcal{G})\;,
\end{align}
where $P$ is the orbifold twist and $Z(\mathcal{G})$ the center of the group
$\mathcal{G}$. In accordance with the usual expectations, all these symmetries 
are gauged, i.e.\ local. 

\subsection*{Acknowledgments}

We would like to thank K.S.~Babu for useful discussions. 
This work is supported by the Deutsche Forschungsgemeinschaft (SFB1258). 
The work of M.R.\ is supported by NSF Grant No.\ PHY-1719438.

\appendix

\section{Torus compactification and symmetries}
\label{app:TorusSymmetries}

In six--dimensions we assume a Yang--Mills theory with upstairs gauge 
group $\mathcal{G}$. Then, the standard Lagrangean for the associated gauge 
bosons $V^M(x,y)$, $M=0,\ldots,5$, reads
\begin{equation}
\mathscr{L} ~=~ -\frac{1}{2}\,\mathrm{tr}\left(F_{MN} F^{MN}\right)\;,
\end{equation}
where $F_{MN}$ denotes the field strength tensor. We expand $V^M(x,y)$ in terms of the generators 
of the Lie algebra of $\mathcal{G}$ in the Cartan--Weyl basis, i.e.\
\begin{equation}
V^M(x,y) ~=~ \sum_I V^M_I(x,y)\,H_I + \sum_{w \in W} V^M_w(x,y)\,E_w ~=~ \sum_a V^M_a(x,y)\,\TCW{a}\;,
\end{equation}
where the index $I$ runs over all Cartan generators $H_I$, $W$ denotes the set
of non--trivial  roots of $\mathcal{G}$ and we denote all Cartan--Weyl
generators collectively by $\TCW{a}$.

An orbifold compactification of this model can be thought of as two steps: 
first we compactify two dimensions on a two--torus $\mathbbm{T}^2$ with 
coordinates  $y=(y_1,y_2)^T$ and then (as described in \Cref{sec:OrbifoldBasics}) on a 
$\mathbbm{T}^2/\Z{N}$ orbifold. To do so, we split the gauge fields $V^M(x,y)$
into components with index $M=\mu$ in  Minkowski space--time and with index
$M=4,5$ in the internal two--torus. From a four--dimensional perspective, the
fields
\begin{equation}
V^\mu \quad\mathrm{and}\quad \chi ~=~ \frac{1}{\sqrt{2}}\left( V^4 + \I\, V^5\right)
\end{equation}
give rise to the gauge bosons of $\mathcal{G}$ and complex scalars,
respectively,  both transforming in the adjoint of $\mathcal{G}$.

\paragraph{Torus compactification.}
We impose boundary conditions on the fields $V^\mu_a(x,y)$ and $\chi_a(x,y)$
compactified on  a two--torus $\mathbbm{T}^2$. To do so, we choose two linearly
independent lattice vectors $e_1$  and $e_2$ that span the torus--lattice.
Depending on the orbifold, we will choose different torus  metrics $G_{ij} = e_i
\cdot e_j$. We take a general, integral linear combination  $n_i e_i$ for $n_i
\in\Z{}$, where summation over $i=1,2$ is implied. Torus periodicity implies
that for all $n_i \in\Z{}$
\begin{subequations}\label{eq:BoundaryConditionsTorus}
\begin{align}
V^\mu_a(x,y + n_i e_i) &~=~ V^\mu_a(x,y)\;, \\
\chi_a(x,y + n_i e_i)  &~=~ \chi_a(x,y)\;.
\end{align}
\end{subequations}
This choice of boundary conditions corresponds to the case of a torus with
trivial gauge background fields (i.e.\ without Wilson lines). Since they are
periodic in $y$, the usual Kaluza--Klein  reduction yields massless modes for
both $V^\mu_a(x,y)$ and $\chi_a(x,y)$ from the four--dimensional point of view.
Consequently, the upstairs gauge symmetry $\mathcal{G}$ remains
unbroken after torus compactification, i.e.\
\begin{subequations}\label{eq:GaugeTransformation}
\begin{align}
V^\mu &~\xmapsto{~\mathcal{G}~} ~ U\,V^\mu\,U^{-1} - \frac{\I}{g}\,\left(\partial^\mu\,U\right)\,U^{-1} \;, \\
\chi  &~\xmapsto{~\mathcal{G}~} ~ U\,\chi\,U^{-1}\;,
\end{align}
\end{subequations}
with $U = U(x)$ in the fundamental representation of $\mathcal{G}$ and $g$
denoting the associated gauge coupling.

\section[D-parity in Pati-Salam from orbifolding]{\boldmath $D$--parity in Pati--Salam from orbifolding\unboldmath}
\label{app:LRParity}

In this appendix, we give an explicit example how one can compute the
action of a residual  symmetry transformation on the unbroken gauge symmetry. To
do so, we consider  $D$--parity from the Pati-Salam example \Cref{sec:PatiSalam}
and work out the  consequences of this \Z2 on $\SO4$. The \so4 algebra is
generated by six  antisymmetric matrices that fulfill
\begin{equation}
\left[M_i,M_j\right]~=~\I\,\varepsilon_{ijk}\,M_k\;,\quad
\left[N_i,N_j\right]~=~\I\,\varepsilon_{ijk}\,M_k\;,\quad \left[M_i,N_j\right]~=~\I\,\varepsilon_{ijk}\,N_k\;.
\end{equation}
An explicit representation can be chosen as
\begin{subequations}
\begin{align}
M_1~=~
\begin{pmatrix}
0&0&0&0\\
0&0&-\I&0\\
0&\I&0&0\\
0&0&0&0
\end{pmatrix}\;,\;\;
M_2~=~
\begin{pmatrix}
0&0&\I&0\\
0&0&0&0\\
-\I&0&0&0\\
0&0&0&0
\end{pmatrix}\;,\;\;
M_3~=~
\begin{pmatrix}
0&-\I&0&0\\
\I&0&0&0\\
0&0&0&0\\
0&0&0&0
\end{pmatrix}\;,\\
N_1~=~
\begin{pmatrix}
0&0&0&-\I\\
0&0&0&0\\
0&0&0&0\\
\I&0&0&0
\end{pmatrix}\;,\;\;
N_2~=~
\begin{pmatrix}
0&0&0&0\\
0&0&0&-\I\\
0&0&0&0\\
0&\I&0&0
\end{pmatrix}\;,\;\;
N_3~=~
\begin{pmatrix}
0&0&0&0\\
0&0&0&0\\
0&0&0&-\I\\
0&0&\I&0
\end{pmatrix}\;.
\end{align}
\end{subequations}
These generators can be ``disentangled'' by making a basis change
$W^{\pm}_i := \frac{1}{2}\left(M_i\pm N_i\right)$, for $i=1,2,3$, 
such that we arrive at the relations
\begin{equation}
\left[W^{+}_i,W^{+}_j\right]~=~\I\,\varepsilon_{ijk}\,W^{+}_k\;, \quad \left[W^{-}_i,W^{-}_j\right]~=~\I\,\varepsilon_{ijk}\,W^{-}_k\;, \quad\left[W^{+}_i,W^{-}_j\right]~=~0\;.
\end{equation}
Hence, we have separated the \so4 into $\su2_\mathrm{L}\oplus\su2_\mathrm{R}$. Now, we take
$U_{\Z2} = \diag(1,1,1,-1)$, see \Cref{eq:PSDefinitionOfZ2}. Following the
diagram~\eqref{eqn:UnbrokenDiscreteSymmetries}, an explicit calculation
reveals that a discrete gauge transformation with $U_{\Z2}$ acts as
\begin{equation}\label{eq:InterchangeSU2s}
W^{+}_i ~\mapsto~ U_{\Z2}\,W^{+}_i\,U_{\Z2}^{-1}~=~ W^{-}_i\;,\quad W^{-}_i ~\mapsto~ U_{\Z2}\,W^{-}_i\,U_{\Z2}^{-1}~=~ W^{+}_i\;.
\end{equation}
Hence, we see explicitly that $U_{\Z2}$ interchanges $\su2_\mathrm{L}$ and $\su2_\mathrm{R}$.

\bibliography{Orbifold}
\bibliographystyle{NewArXiv}
\end{document}